\documentstyle[aps,prl,epsfig]{revtex}

\begin{document}

\draft

\twocolumn[\hsize\textwidth\columnwidth\hsize\csname
@twocolumnfalse\endcsname

\title{Charge localization and phonon spectra in hole doped La$_{2}$NiO$_{4}$}

\author{R. J. McQueeney, A. R. Bishop, and Ya-Sha Yi}
\address{Los Alamos National Laboratory, Los Alamos, New Mexico 87545}

\author{Z. G. Yu}
\address{Department of Chemistry, Iowa State University, Ames, Iowa 50011}

\date{Received on January 13, 2000}

\maketitle

\begin{abstract}
The in-plane oxygen vibrations in La$_{2}$NiO$_{4}$ are investigated for 
several hole-doping concentrations both theoretically and experimentally via 
inelastic neutron scattering.  Using an inhomogeneous Hartree-Fock plus RPA 
numerical method in a two-dimensional Peierls-Hubbard model, it is
found that the doping induces stripe ordering of localized charges,
and that the strong electron-lattice coupling causes the in-plane 
oxygen modes to split into two subbands. This result
agrees with the phonon band splitting observed by inelastic neutron 
scattering in La$_{2-x}$Sr$_{x}$NiO$_{4}$.
Predictions of strong electron-lattice coupling in La$_{2}$NiO$_{4}$,
the proximity of both oxygen-centered and nickel-centered charge
ordering, and the relation between charged stripe ordering and the
splitting of the in-plane phonon band upon doping are emphasized. 
\end{abstract}
\smallskip
\pacs{PACS numbers: 75.60.Ch, 74.25.Kc, 71.45.Lr, 71.38.+i}
]

There is currently great interest in the importance of charge
localization and ordering tendencies in a variety of doped transition
metal oxides: including nickelates, bismuthates, cuprates, and
manganites\cite{tranquada94,tranquada96,chen93,cheong9497,nakajima97,tranquada9597,ramirez97}.
Recent experiments have suggested nanoscale coexistence of charge and
spin ordering, as well as related multiscale dynamics
\cite{tranquada94,tranquada96,chen93,cheong9497,nakajima97,tranquada9597,ramirez97,blumberg98,katsufuji96,yamamoto98}.
The cuprates have been widely investigated, both theoretically and
experimentally, as this inhomogeneity may be related to
high-temperature
superconductivity\cite{theoryblock,emery93,emery95,castroneto96,white97,krotov97,zaanen96,bianconi96}.

The nickelates are considered strong electron-lattice (e-l) coupling
systems, which helps stabilize charge-ordering in the form of "stripe"
phases \cite{wochner98,zaanen94}.  For the commensurate 1/3 doping
case of La$_{1.67}$Sr$_{0.33}$NiO$_{4}$, it has been shown in optical
absorption and Raman scattering experiments that new phonon modes
appear when the temperature is lowered below the stripe-ordering
temperature ($T_{so}$=240 K); this is a signature of the stripe
formation \cite{blumberg98,katsufuji96,yamamoto98}.  Until now, only
the temperature dependence and the apical oxygen (Ni-O(2)) vibrations
have been investigated.  However, the doping dependence and the
in-plane oxygen vibrations (Ni-O(1) stretching modes) are also very
important for the properties of the quasi-two dimensional nickelate
materials.  In this study, we use an inhomogeneous Hartree-Fock (HF)
plus random-phase approximation (RPA) numerical method for a
two-dimensional (2D), four-band Peierls-Hubbard model, to interpret
the inelastic neutron scattering spectra. This reveals specific
signatures of the stripe patterns in the in-plane oxygen phonons.

Our main results are: (i) There is agreement between the results from
our multiband model including electron-electron and e-l interactions
and the inelastic neutron scattering spectra for the in-plane oxygen
vibrations with various commensurate hole-doping concentrations; (ii)
The theoretical results predict new vibrational modes (``edge modes'')
which are associated with oxygen motions near localized holes or in
the vicinity of stripes; (iii) The e-l coupling strength, at which the
best agreement between our model and the inelastic neutron scattering
data is achieved, is close to the transition from an oxygen-centered
stripe phase to a nickel-centered one.  This suggests that the
nickelates may be in a mixed state of both stripe phases, and
sensitive to temperature, pressure and magnetic field.

The inelastic neutron scattering spectra were measured on polycrystalline 
La$_{2-x}$Sr$_{x}$NiO$_{4}$ for various doping concentrations, 
$x=$ 0, 1/8, 1/4, 1/3, 1/2.  Time-of-flight neutron
scattering measurements were performed on the Low Resolution Medium
Energy Chopper Spectrometer at Argonne National Laboratory's Intense
Pulsed Neutron Source.  For all measurements, an incident neutron
energy of 120 meV was chosen and data were summed over all scattering
angles from $2^{\circ}-120^{\circ}$.  Detailed information
on the experiment, as well as the preparation of the samples used, can
be found in Ref. \cite{mcqueeney99}.  The focus is on the
(neutron-scattering-weighted) generalized density-of-states (GDOS) of
phonons and particular attention is given to the in-plane oxygen
vibrations, i.e., the Ni-O(1) stretching modes.  

\begin{figure}
\centering
\epsfxsize\columnwidth
\epsffile{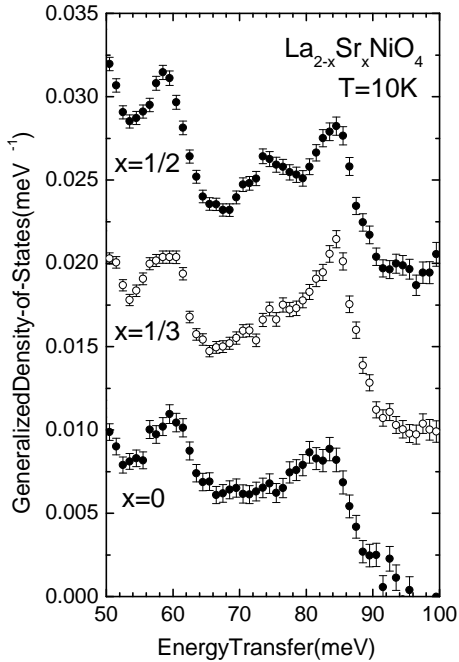}
\caption{The generalized phonon density-of-states for three
concentrations of La$_{2-x}$Sr$_{x}$NiO$_{4}$.  The frequency region
shown (above 65 meV) consists of in-plane polarized oxygen modes
(breathing modes) which are well-separated from other types of
phonons.  Data are offset vertically for clarity.}
\label{fig1}
\end{figure}

Figure \ref{fig1} shows the experimental GDOS for several hole
concentrations at $T$=10 K for phonon modes in the range from 50-100
meV\@.  From the analysis of lattice dynamical shell model
calculations, it is known that the in-plane Ni-O(1) oxygen stretching
modes are separated in frequency from other vibrations, and the phonon
intensity above 65 meV is associated entirely with these vibrations.
For the low doping samples (\(x= 1/8\) and \(1/4\), not shown), there
is little change in the $\sim$81 meV phonon band \cite{mcqueeney99}, but
for the \(x=1/3\) and \(x=1/2\) samples, a new peak appears around 75
meV along with a slight hardening of the main band\@.  This feature is
interpreted as a splitting of the 81 meV Ni-O(1) stretching band into
two phonon bands centered at approximately 83 meV and 75 meV.

In order understand this dependence of the Ni-O(1) stretching modes on
hole concentration and its possible reflection of stripe ordering, we
have performed a calculation of the phonon spectrum in a minimal
Peierls-Hubbard model in 2D\@. Due to the strong e-l coupling expected
in the nickelates, we resort to modeling with an inhomogeneous HF plus
RPA numerical approach \cite{batistic92,yonemitsu92,yonemitsu93}.
This has proven to be a very robust method for studying charge
localization and stripe formation, especially when electron-lattice
coupling is strong, obviating subtle many-body effects and quantum
fluctuations \cite{yi98}.

We use a 2D four-band extended Peierls-Hubbard model of a doped
NiO$_{2}$ plane, which includes both electron-electron and e-l
interactions \cite{yonemitsu92,yonemitsu93}.  Here, for nickelate
oxides, besides the \(d_{x^{2}-y^{2}}\) orbital used in the cuprate
oxide models \cite{yonemitsu92}, the \(d_{3z^{2}-1}\) Ni $d$ orbitals
must be included to account for the higher spin state (\(S=1\)) at
half-filling (i.e. undoped).  Our model Hamiltonian is \cite{yi98}:

\begin{eqnarray}
H & = & \sum_{\langle ij
\rangle,m,n,\sigma}t_{im,jn}(u_{ij})(c^{\dagger}_{im\sigma}
c_{jn\sigma}+{\rm H.c.})\nonumber \\
& + & \sum_{i,m,\sigma}\epsilon_m
c^{\dagger}_{im\sigma} c_{im\sigma}+\sum_{\langle ij
\rangle}\frac{1}{2}K_{ij}u^2_{ij}+H_c,
\end{eqnarray}
where $c^{\dagger}_{im\sigma}$ creates a hole with spin $\sigma$ at
site $i$ in orbital $m$ (Ni $d_{x^2-y^2}$, $d_{3z^2-1}$, or O $p$).
The Ni-O hopping $t_{im,jn}$ has two values: $t_{pd}$ between
$d_{x^2-y^2}$ and $p$ and $\pm t_{pd}/\sqrt{3}$ between $d_{3z^2-1}$
and $p$.  The O-site electronic energy is \(\epsilon_{p}\), and
Ni-site energies are \(\epsilon_{d}\) and \(\epsilon_{d}+E_{z}\) for
\(\epsilon_{m}\), with \(E_{z}\) the crystal-field splitting on the Ni
site.  $H_c$ describes the electron correlations in the Ni orbitals,

\begin{eqnarray}
H_c & = &\sum_{im}(U+2J)n_{im\uparrow}n_{im\downarrow}
-\sum_{i,m\ne n} 2J {\rm\bf S}_{im}\cdot {\rm\bf S}_{in}\nonumber\\ 
& + &\sum_{i,m\ne n,\sigma,\sigma^{\prime}}(U-J/2)n_{im\sigma}
n_{in\sigma^{\prime}}\nonumber\\
& + &\sum_{i,m,n}Jc^{\dagger}_{im\uparrow}
c^{\dagger}_{im\downarrow}c_{in\downarrow}c_{in\uparrow}
\end{eqnarray}  
The electron-electron interactions include the on-site Ni Coulomb
repulsions ($U$) as well as the Hund interaction ($J$) at the same Ni
site to account for the high spin situation.  (The interplay of the
two orbitals can also lead to pseudo Jahn-Teller distortions, but
these are not our focus here).  We emphasize that, due to the large
spin at the nickel site, Hund's rule leads to ferromagnetic exchange
coupling $-2J$, and \({\bf
S_{im}}=\frac{1}{2}\sum_{\tau\tau'}c_{im\tau}^{\dagger}
\mbox{\boldmath $\sigma_{\tau\tau'}$}c_{im,\tau'}\),with
\(\mbox{\boldmath $\sigma$}\) the Pauli matrix.  For the e-l
interaction, we consider that the Ni-O hopping is modified linearly by
the O-ion displacement $u_{ij}=u_{\rm O}$ as
$t_{im,jn}(u_{ij})=t_{im,jn}(1\pm \alpha u_{\rm O})$, where the $+$
($-$) applies if the bond shrinks (stretches) with positive $u_{\rm
O}$.  For the lattice terms, we study only the motion of O ions along
the Ni-O bonds---other oxygen (or Ni) distortion modes can readily be
included if necessary.  It is known that for the nickelate oxides the
e-l coupling is stronger than in cuprate oxides
\cite{wochner98,zaanen94}, and is therefore likely to play an even
more decisive role in the formation, localization, and nature of
stripe phases.  We adopt the following representative paremeters for
the nickelate materials \cite{zaanen94}: \(t_{pd}=1\),
\(\Delta=\epsilon_{p}-\epsilon_{d}=9\), \(U=4\), \(J=1\), \(E_{z}=1\)
and \(K=32t_{pd}/\AA^{2}\) (all in units of \(t_{pd}\)).  In real
oxides \cite{zaanen94,vanelp92}, \(t_{pd}\) is estimated to be in the
range 1.3 eV $\sim$ 1.5 eV\@.  The electron-lattice coupling strength
is varied to achieve a best fit to the neutron scattering data; we
find \(\alpha \approx 3.0\).  The commensurate doping cases are
examined in a 4x4 unit supercell for \(x=0\), 1/2 and a 3x3 unit supercell
for \(x=1/3\).  Periodic boundary conditions are used.

\begin{figure}
\centering
\epsfxsize\columnwidth
\epsffile{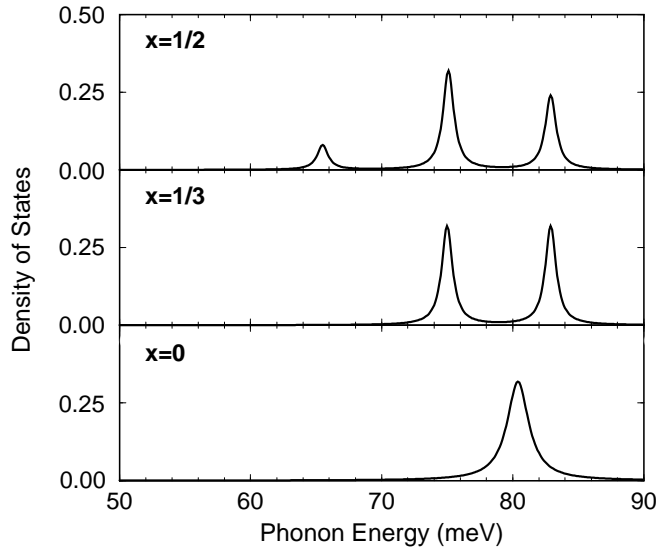}
\caption{The calculated densities-of-states of the oxygen breathing
modes for various doping concentrations.  The results have been
broadened with a lorentzian of width 2 meV for $x=0$ and 1 meV for
$x=1/3$ and 1/2.  The electron-lattice coupling constant used
is \(\alpha=3.0\).}
\label{fig2}
\end{figure}

The densities-of-states (DOS) of in-plane phonons were calculated from
our model at $x=0$, 1/3, 1/2 and are shown in Fig. \ref{fig2}. For the
undoped case, as the ground state is spatially homogeneous, only one
oxygen phonon band appears, centered around 80.5 meV\@.  When holes
are added into the NiO$_{2}$ plane at \(x=1/3\), the ground state is
found to be a stripe pattern with more holes accumulating along the
(1,1,0) direction, forming an antiphase domain wall within the
original antiferromagnetic background.  This is consistent with many
neutron, optical and Raman scattering experiments
\cite{cheong9497,blumberg98,katsufuji96,yamamoto98}.  Interestingly,
we find that a new phonon band appears centered at 75 meV\@.  In
addition, there is a hardening of the main phonon band (to 83 meV)
corresponding to an overall splitting of 8 meV\@.  From examination of
the eigenvectors, the main character of this band is local oxygen
vibrations in the vicinity of the stripe, i.e. having the nature of
localized "edge modes" (see \cite{yi98} for more details).  For the
higher doping \(x=1/2\), the charge-ordering takes on a commensurate
checker-board pattern.  A similar splitting of the 85 meV phonon band
into two phonon bands around 83 and 75 meV is again found.  At this
half doping, the checker-board ground state is in essence a
commensurate charge-density-wave (CDW) system with the nature of an
ordered binary alloy.

The above results are in agreement with the GDOS data obtained from
inelastic neutron scattering, which are shown in Fig. \ref{fig1}.  In
addition to the splitting energy, even the slight hardening of the
main band observed experimentally is accounted for in the model.
Besides the new phonon modes centered at 75 meV appearing for the 1/3
and 1/2 doping, another low intensity phonon mode around 65 meV is
also predicted in our model at \(x=1/2\) (Fig. \ref{fig2}). The
signature for these modes are weak, however, so that they may be
difficult to detect in the current experiment.  In so far as we have
included only a small subset of the possible oxygen displacement
patterns and wavevectors in the model (whereas the neutron scattering
experiment samples all wavevectors and polarizations), the relative
intensities and widths of the bands obtained by experiment and theory
cannot be usefully compared.

\begin{figure}
\centering
\epsfxsize\columnwidth
\epsffile{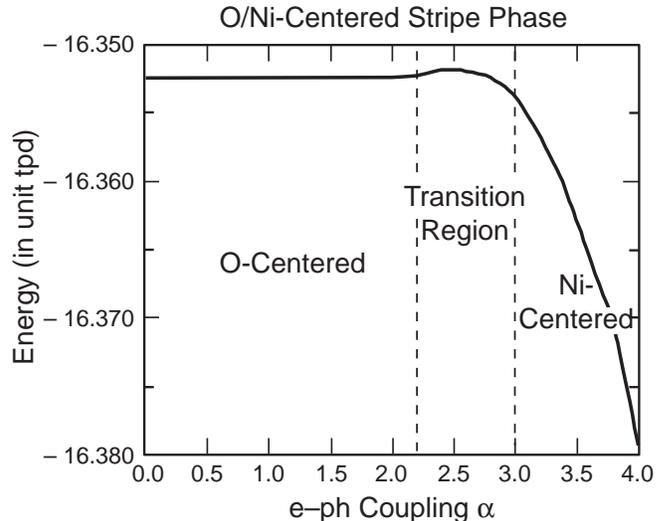}
\caption{The energy dependence on the electron-lattice coupling
$\alpha$ for the 1/3 doped nickelates (solid line).  For 
\(\alpha \leq 2.2\), the ground state is in an O-centered stripe
phase.  While for \(\alpha \geq 3.0\), the Ni-centered state is found as the
ground state.  The sensitive transition region is from \(\alpha=2.2\)
to \(\alpha=3.0\).}
\label{fig3}
\end{figure}

We emphasize that the excellent agreement between our model and the
GDOS experimental data is achieved by varying the e-l coupling
strength to match the positions of the phonon bands.  As noted, the
choice of \(\alpha \approx 3\) best fits the data; our model
calculations predict a variation of the phonon splitting with $\alpha$
which is quite large (\(\Delta\omega_{split}/\Delta\alpha \approx\) 7
meV).  Most strikingly, as illustrated in Fig. \ref{fig3}, an
O-centered stripe is found as the ground state at small
\(\alpha \alt 2.2\), while for a larger \(\alpha \agt
3.0\), a Ni-centered stripe is found as the ground state\cite{yi98}.
The transition region includes \(\alpha \approx 3.0\), where the best
agreement between Fig. \ref{fig2} based on our model and
Fig. \ref{fig1} on the inelastic neutron scattering spectra is
achieved.  It has been suggested from various experimental data that
stripe formation for $x=1/3$ can not be simply assigned as Ni-centered
or O-centered \cite{wochner98}, but is also dependent on temperature.
Our comparison of theory and experiment provides a possible
explanation on the sensitivity of stripe formation; they suggest that
La$_{1.67}$Sr$_{0.33}$NiO$_{4}$ may be in a mixed stripe phase state,
and also in a region of sensitivity to temperature, pressure, magnetic
field, etc.

In conclusion, we have made a study of oxygen breathing lattice
vibrations in La$_{2-x}$Sr$_{x}$NiO$_{4}$ via inelastic neutron
scattering compared with predictions of a 2D four-band model,
including both electron-lattice and electron-electron interactions.
The in-plane oxygen vibrations above 65 meV were thoroughly
investigated.  The splitting of the in-plane 81 meV band upon doping
into two subbands centered around 75 meV and 83 meV is observed
experimentally and predicted theoretically, and interpreted in terms
of new localized phonon modes (``edge modes'' at charge localized
stripes).  The excellent agreement between the experiment and the
model strongly supports the view that strong electron-lattice coupling
in this kind of material plays a decisive role on the charge
localization and mesoscopic stripe formation.  For the doping at
$x=1/3$, at which stripes are found both in experiments and our model,
our results suggest that there may be a mixed state of O- and
Ni-centered stripe phases, and sensitivity to temperature, pressure,
and magnetic field.  Our model calculations also predict distinctive
dispersion of the phonon bands, as well as inhomogeneous
magnetoelastic coupling along the boundaries between charge-rich and
magnetic nanophase domains \cite{yi98}.  These predictions require
additional experiments for their confirmation and consequences.

We have benefitted from valuable discussions with
Dr. J. T. Gammel. This work is supported (in part) by the U.~S.\
Department of Energy under contract W-7405-Eng-36 with the
University of California.  This work has benefited from the use of the
Intense Pulsed Neutron Source at Argonne National Laboratory.  This
facility is funded by the U.~S.\ Department of Energy, BES-Materials
Science, under contract W-31-109-Eng-38.


\end{document}